# Study of a New Neuron

*Stephen L. Adler and Gyan V. Bhanot*

Institute for Advanced Study

Princeton, NJ 08540

*John D. Weckel*

Physics Department, Princeton University

Princeton, NJ 08544

## ABSTRACT

We study a modular neuron alternative to the McCulloch-Pitts neuron that arises naturally in analog devices in which the neuron inputs are represented as coherent oscillatory wave signals. Although the modular neuron can compute $XOR$ at the one neuron level, it is still characterized by the same Vapnik-Chervonenkis dimension as the standard neuron. We give the formulas needed for constructing networks using the new neuron and training them using back-propagation. A numerical study of the modular neuron on two data sets is presented, which demonstrates that the new neuron performs at least as well as the standard neuron.

## 1. Introduction

The standard basic computational element for neural networks is the McCulloch-Pitts (1943) neuron, given by

$$v_j^{out} = \theta \left( \sum_{k=1}^{N} w_{jk} v_k^{in} - T_j \right), \tag{1a}$$

in which $v_k^{in}$ and $v_j^{out}$ are respectively inputs and outputs which take the values 0 and 1, $w_{jk}$ and $T_j$ are respectively real number weights and a threshold, and $\theta$ is the Heavside step function defined by

$$\theta(x) = \begin{cases} 1 & x \geq 0 \\ 0 & x < 0 \end{cases}. \tag{1b}$$

Networks constructed from the neuron of Eqs. (1a,b) (which we refer to henceforth as the *standard or basic neuron*) have been extensively studied (see, e.g., Hertz, Krogh, and Palmer, 1991). Over the years, with the aim of at taining greater computational power or greater biological realism, neurons more complicated than that of Eqs. (1a,b) have been discussed. These include neurons with complex number weights (Denker, 1986; Kim and Guest, 1990), neurons with multiple internal states (Valiant, 1994), and neurons with thresholds and other parameters which respond adaptively to the inputs (Abbott, 1994 and Valiant, 1994). In this paper we analyze in detail a specific generalization of the neuron of Eqs. (1a,b) which we term the *modular neuron* (Adler, 1991; Adler, 1993), given by

$$v_j^{out} = \theta \left( t_j - \left| \sum_{k=1}^{N} w_{jk} v_k^{in} - u_j \right| \right). \tag{2a}$$

Here $|x|$ is the modulus function defined for real $x$ by

$$|x| = \begin{cases} x & x \geq 0 \\ -x & x < 0 \end{cases}, \tag{2b}$$

with the generalization to complex $x$ given by

$$|x| = \left[(Re\, x)^2 + (Im\, x)^2\right]^{1/2} \; ; \tag{2c}$$

the weights $w_{jk}$ and offset $u_j$ can be either real or complex number quantities, while the threshold $t_j$ is real and positive. In the following sections we analyze theoretical aspects of the modular neuron with real number weights, and give numerical comparisons of networks constructed from standard and modular neurons. In these studies, we also include in the comparisons cases in which the neurons in the output layer of the network are unthresholded *linear neurons*, defined by

$$v_j^{out} = \sum_{k=1}^{N} w_{jk} v_k^{in} - T_j \; . \tag{3}$$

The specific functional form of Eqs. (2a,b) is motivated by the consideration of analog devices in which the neuron inputs are represented by phase-coherent oscillatory waves. In an electronic neuron proposed by Adler (1993), neuron inputs and the offset are represented by phase-coherent alternating current signals of amplitude $v_k^{in}$ and $u_j$ respectively, with the synapses obtained by capacitive coupling of these signals to a summation line. The alternating current signal on the summation line is then rectified and filtered, producing a direct current signal proportional to its modulus. This signal is then compared with a threshold direct current signal, with the resultant used to gate an alternating current source which is the neuron input to the next layer, as illustrated in Fig. 1(a) for the case in which all synapses are excitatory. Synapses which are either excitatory or inhibitory, corresponding to positive or negative weights $w_{jk}$, can be produced in this architecture by a two phase input as in Fig. 1(b), while complex number weights can be produced by using a three phase input, as shown in Fig. 1(c). An optical implementation of neural nets, in which the weights are realized as a volume hologram in a photorefractive crystal, has been discussed by Psaltis, Brady and Wigner (1988). Both incoherent and coherent implementations are possible; in the latter, the neural output intensity is proportional to

$$I_j = \left| \sum_{k=1}^{N} w_{jk} v_k^{in} \right|^2 ,  \qquad (4)$$

with $v_k^{in}$ complex wave amplitudes and with $w_{jk}$ complex weightings induced by the volume hologram. Electronic processing of $I_j$ by threshold comparison and gating of a neuron output signal then leads to a neuron with the modular neuron form of Eqs. (2a,c).

Before turning to a detailed discussion of properties of the modular neuron, let us address a general argument which is often directed against considering neurons more complicated than the standard neuron. This argument starts from the observation that the standard neuron, when used in networks, can compute any function (as is discussed in the following section). Therefore, the reasoning goes, it is not necessary to use neurons more complicated than the standard one, and if one does go in the direction of increased neuron complexity, then why not go the full length and make each neuron a complete digital computer chip? This argument, we believe, ignores some basic systems considerations for highly interconnected neural networks, in which one ultimately hopes to make the number $N$ of input synapses per neuron very large. (In the human brain, $N$ is typically of order $10^4$; see, e.g. Churchland and Sejnowski, 1992.) Given a "black box" neuron of unknown function, with $N \gg 1$ input synapses and one output, two hypotheses would seem *a priori* reasonable: (i) Since the synapse multiplicity is high, the synapse mechanism should be simple and universal. (ii) The complexity of the neural processing mechanism feeding the single output should be less than or of order the *total* input synapse complexity,

$$\text{neural processing complexity} \quad \lesssim \quad N \quad \times \quad \text{input synapse complexity} . \quad (5)$$

The first of these hypotheses is satisfied (with minor exceptions) in the human brain. The second permits neurons with considerably greater processing complexity than the minimal standard neuron, but is not compatible with a neuron as complex as a digital computer chip; it also appears to be biologically plausible, in view of the fact that the brain contains tens to hundreds of different neuron cell types. Applying these systems considerations to artificial neuron construction argues that the primary emphasis should be on achieving a simple and robust realization of the synapses, even if (as in the case of coherent wave synapse realizations) this requires a more complex neuron processor than that of the standard neuron.

## 2. Theoretical aspects of the modular neuron

We discuss here a number of aspects of the modular neuron of Eqs. (2a,b) which are relevant for neural network applications, considering only the case in which the weights $w_{jk}$ and offset $u_j$ are real numbers (which can be either positive or negative). As in analyzing properties of the standard neuron, it is convenient to represent the $N$ neuron inputs $v_1^{in}, \ldots, v_N^{in}$ as a vector $\vec{v}^{in}$ in an $N$-dimensional vector space. In geometric terms, the standard neuron of Eqs. (1a,b) selects for a half-space

$$T_j \leq \vec{w}_j \cdot \vec{v}^{in} , \qquad (6a)$$

with $\vec{w}_j = (w_{j1}, \ldots, w_{jN})$ the $N$-dimensional vector formed from the weights $w_{jk}$. In similar terms, the modular neuron of Eqs. (2a,b) selects for the interior of a "sandwich" of thickness $2t_j$ bounded by two parallel hyperplanes,

$$-t_j + u_j \leq \vec{w}_j \cdot \vec{v}^{in} \leq t_j + u_j . \qquad (6b)$$

If we let $u_j = T_j + t_j$ and take the limit $t_j \to \infty$ with $T_j$ fixed, then the interior domain of Eq. (6b) approaches the half-space domain of Eq. (6a). Therefore the modular neuron contains the standard neuron as a limiting case, which implies that the modular neuron is at least as powerful computationally as the standard neuron.

In some problems of practical importance, the modular neuron has greater computational power than the standard one. To see this, let us take $N = 2$ and consider the "exclusive or" ($XOR$) and "exclusive nor" ($XNOR$) functions with $v^{out}$ given by the truth tables

$$
\begin{array}{cccc|cccc}
XOR: & v^{out} & v_1^{in} & v_2^{in} & XNOR: & v^{out} & v_1^{in} & v_2^{in} \\
 & 0 & 0 & 0 & & 1 & 0 & 0 \\
 & 1 & 0 & 1 & & 0 & 0 & 1 \\
 & 1 & 1 & 0 & & 0 & 1 & 0 \\
 & 0 & 1 & 1 & & 1 & 1 & 1
\end{array}
\tag{7a}
$$

A well-known result (Minsky and Papert, 1969) of perceptron theory states that a single standard neuron cannot represent either $XOR$ or $XNOR$; a two-layer network constructed from standard neurons is needed to do this. On the other hand, the modular neuron

$$v^{out} = \theta\left(\frac{1}{2} - |v_1 + v_2 - 1|\right) \tag{7b}$$

represents $XOR$, while the modular neuron

$$v^{out} = \theta\left(\frac{1}{2} - |v_1 - v_2|\right) \tag{7c}$$

represents $XNOR$. In addition to the modular neuron of Eqs. (2a,b), one can introduce a complementary modular neuron in which the sign of the argument of the step function is reversed, according to

$$v_j^{out} = 1 - \theta\left(t_j - \left|\sum_{k=1}^{N} w_{jk} v_k^{in} - u_j\right|\right) = \theta\left(\left|\sum_{k=1}^{N} w_{jk} v_k^{in} - u_j\right| - t_j\right). \tag{8a}$$

The complementary neuron evidently selects for the exterior of a sandwich of thickness $2t_j$

bounded by two parallel hyperplanes,

$$\vec{w}_j \cdot \vec{v}^{in} \leq -t_j + u_j \quad \text{or} \quad \vec{w}_j \cdot \vec{v}^{in} \geq t_j + u_j \ . \tag{8b}$$

Using the complementary form of the modular neuron, we can represent $XOR$ as

$$v^{out} = \theta \left( |v_1 - v_2| - \frac{1}{2} \right) , \tag{8c}$$

and $XNOR$ as

$$v^{out} = \theta \left( |v_1 + v_2 - 1| - \frac{1}{2} \right) . \tag{8d}$$

Having a neuron of greater computational power is not an automatic advantage in constructing neural nets, because the more flexible one makes a net in fitting an arbitrary function, the less well it can be expected to generalize. A quantitative measure of the ability of a neuron (or a network) to generalize is provided by the Vapnik-Chervonenkis (VC) dimension $d_{VC}$ (Vapnik and Chervonenkis, 1971; Abu-Mostafa, 1989), which in $N$ dimensions is defined as the cardinality of the largest set for which the neuron (or network) generates all possible dichotomies. According to a standard result, the $VC$ dimension of the standard neuron is

$$d_{VC}(\text{standard neuron}) = N + 1 \ . \tag{9}$$

(Some authors define $d_{VC}$ as the cardinality of the *smallest* set for which the neuron *cannot* generate all possible dichotomies, which is larger by 1 than the definition used here.) Since the modular neuron of Eqs. (2a,b) contains the standard one as a limiting case, it can generate any dichotomy which can be generated by a standard neuron, which implies that

$$d_{VC}(\text{modular neuron}) \geq N + 1 \ . \tag{10a}$$

On the other hand, let us now construct a point set of cardinality $N + 2$ in $N$ dimensions, which has a dichotomy which *cannot* be generated by the modular neuron. To do this, take any set consisting of $N + 1$ points at the vertices of a convex polyhedron, and a final point lying interior to the polyhedron. The dichotomy of this set, in which the $N + 1$ points at the vertices are assigned the value 1 whereas the one interior point is assigned the value 0, cannot be produced by the modular neuron of Eqs. (2a,b), because any sandwich containing the $N + 1$ exterior points must, by convexity of the polyhedron, contain the interior point as well. This example implies that

$$d_{VC}(\text{modular neuron}) < N + 2 , \qquad (10b)$$

which with Eq. (10a) implies that

$$d_{VC}(\text{modular neuron}) = N + 1 . \qquad (10c)$$

Hence *despite the fact that it can compute $XOR$ and $XNOR$, the modular neuron has the same Vapnik-Chervonenkis dimension, and therefore is in the same generalization class, as the standard neuron.* In Figs. 2(a)–(c), we show in $N = 2$ the dichotomies of four-point sets corresponding to $XOR$ and $XNOR$ that are generated by the modular neuron, as well as a four-point set and dichotomy constructed as above that cannot be generated by the modular neuron.

So far we have discussed single neuron aspects of the modular neuron; let us now turn to considerations which apply when modular neurons are used to build neural networks. In constructing modular neural networks, we shall always take the network outputs to be *linear* *functions* of the outputs of the final layer neurons,

$$O_j = \lambda_j \theta \left( t_j - \left| \sum_{k=1}^{N_{f-1}} w_{jk} v_k^{in} - u_j \right| \right) + \kappa_j , \qquad j = 1, \ldots, N_f , \qquad (11a)$$

where $N_f$ and $N_{f-1}$ are the numbers of neurons in the final and semifinal layers respectively, and where for each $j$ the quantities $\lambda_j$ and $\kappa_j$ are treated as free parameters along with the neuron parameters $w_{jk}, u_j, t_j$. This has two advantages. First of all, in function fitting applications, it eliminates the necessity to do a preliminary shifting and rescaling of the target output values, since this is automatically accomplished by the optimization with respect to the parameters $\lambda_j$ and $\kappa_j$. Secondly, since the general form of Eq. (11a) is invariant under the substitution $\theta \to 1 - \theta$ together with appropriate redefinitions of the $\lambda_j$ and $\kappa_j$ values, it is not necessary to consider as separate cases networks in which some of the final layer neurons have the complementary modular neuron form of Eq. (8a). This conclusion is also automatically true for layers below the final layer, since the linear expression

$$\sum_k w_{jk} v_k^{in} - u_j \qquad (11b)$$

which appears as the modular neuron argument is already form-invariant under the substitution $v_k^{in} \to 1 - v_k^{in}$ together with appropriate redefinitions of the weights $w_{jk}$ and offset $u_j$. Consequently, we never have to explicitly consider neurons with the complementary modular neuron form of Eq. (8a); this possibility is automatically taken into account by using the modular neuron of Eqs. (2a,b), together with a linear remapping of the outputs as in Eq. (11a). In doing comparisons with standard neurons in the output layer, we again take the network outputs to be linear functions of the neuron outputs, now according to

$$O_j = \lambda_j \theta \left( \sum_{k=1}^{N_{f-1}} w_{jk} v_k^{in} - T_j \right) + \kappa_j . \qquad (11c)$$

When the linear neuron of Eq. (3) is used in the output layer, the additional linear remapping is redundant, and is omitted.

In training modular neural networks by back-propagation, it is necessary to replace the step function by a sigmoidal function, just as in training standard neural networks. Corresponding to the standard neuron replacement

$$\theta(x) \to \frac{1}{1 + e^{-2\beta x}} \equiv g_S(x) , \qquad (12a)$$

we make for the modular neuron the replacement

$$\theta(t - |x|) = \theta(x + t) - \theta(x - t) \to g_S(x + t) - g_S(x - t) \equiv g_M(x, t) ; \qquad (12b)$$

from Eqs. (12a,b), the first derivatives of $g_S(x)$ and $g_M(x,t)$ are readily computed to be

$$\frac{\partial g_S(x)}{\partial x} = 2\beta g_S(x)[1 - g_S(x)] ,$$
$$\frac{\partial g_M(x,t)}{\partial x} = 2\beta \{g_S(x + t)[1 - g_S(x + t)] - g_S(x - t)[1 - g_S(x - t)]\} ,$$
$$\frac{\partial g_M(x,t)}{\partial t} = 2\beta \{g_S(x + t)[1 - g_S(x + t)] + g_S(x - t)[1 - g_S(x - t)]\} . \qquad (12c)$$

It is now straightforward (Hertz, Krogh, and Palmer, 1991, pp. 115-120) to implement the back-propagation algorithm for the modular neuron, using the cost function

$$E = \frac{1}{2} \sum_{j=1}^{N_f} (\zeta_j - O_j)^2 , \qquad (12d)$$

with $\zeta_j$ the target outputs and $O_j$ the linearly remapped network outputs as given by Eq. (11a).

We conclude our theoretical analysis with a brief discussion of function fitting and generalization by a network constructed using modular neurons. Lapedes and Farber (1988) have pointed out that using standard neurons one can represent any function with a two layer network. In the first layer, half lines $L_\ell < x_\ell$ and $U_\ell < x_\ell$ (with $L_\ell < U_\ell$) for any coordinate $x_\ell$ can be selected by standard neurons $\theta(x_\ell - L_\ell)$ and $\theta(x_\ell - U_\ell)$; in the second layer these are combined to give $\theta(x_\ell - L_\ell) - \theta(x_\ell - U_\ell)$, which takes the value 1 on the interval $L_\ell < x_\ell < U_\ell$, and such combinations selecting for intervals in the various coordinates are further combined to select for a cell in the multi-dimensional coordinate space. Each neuron in the second layer then selects for a specific cell in the multi-dimensional coordinate space, and has an output weighted to give the function value appropriate to that cell. The same construction can clearly be implemented using modular neurons in the first layer, followed by either modular or standard neurons in the second layer. Since $\theta(x_\ell - L_\ell) - \theta(x_\ell - U_\ell)$ is just a modular neuron,

$$\theta(x_\ell - L_\ell) - \theta(x_\ell - U_\ell) = \theta\left(\frac{1}{2}(U_\ell - L_\ell) - |x_\ell - \frac{1}{2}(U_\ell + L_\ell)|\right), \tag{13}$$

the modular neuron implementation of the Lapedes-Farber argument uses half as many first layer neurons as does the implementation using standard neurons. This argument, together with the fact that the modular neuron contains the standard one as a limiting case, suggests that modular neurons should perform as well as or better than a standard neuron in function fitting applications, with potential reductions of up to a factor of two in network size. Actual gains achieved will of course depend strongly on the details of the functions being represented and the network architecture employed.

Generalization by neural networks has been analyzed by Baum and Haussler (1989), who use Vapnik-Chervonenkis theory to give sufficient conditions for valid generalization by a network. Baum and Haussler proceed by deriving an upper bound for the $VC$ dimension of a feed forward network, expressed in terms of the $VC$ dimensions of the basic computational elements or neurons. Since we have seen earlier that a modular neuron has the same $VC$ dimension as a standard neuron with the same number of inputs, the bounds given in Sec. 3 of the Baum-Haussler paper carry over directly to networks constructed using modular neurons. That is, given a standard and a modular neural network with the same architecture, the Baum-Haussler bounds will be the same; to the extent that actual generalization is represented by the bounds, generalization in the two cases will be comparable. Since the Baum-Haussler bounds scale with network size, the discussion of function representation given earlier suggests that improved generalization may be achieved by using modular neurons to the extent that one can achieve function representation with reduced network size. However, we did not attempt a systematic study of this issue in the numerical experiments described in the following section.

### 3. Numerical Study

In our numerical study, we used the back-propagation algorithm to compare modular with standard neurons on two data sets, one small and one relatively large, that we found in the literature.

The small data set (Chukwujekwu Okafor, Marcus, and Tipireni, 1991) was a table of 24 values for the surface roughness (SR) and the bore tolerance (BT) as functions of the rms horizontal resultant force (RF), rms spindle acceleration (SA) and acoustic emission (AE) event magnitude, in a circular end milling machine.

The larger data set (Odewahn, Stockwell, Pennington, Humphreys, and Zumach, 1992) came from a star/galaxy discrimination study. Fourteen input attributes were used in this study to determine if the object was a star or a galaxy. We used the large diameter regime data ($146 \mu m < D < 330 \mu m$ - called LP by Odewahn et al.) which had 1698 independent data elements for stars and 1068 data elements for galaxies.

The cost function used for training was the standard one given in Eq. (12d). We made ten training experiments using the data and the results given below represent the average performance of the net over these experiments. For each experiment, $w_{jk}$, $t_j$, $u_j$, and where relevant, $\lambda_j$ and $\kappa_j$, were set initially to random values in the interval $[-1, 1]$. The data was divided so that half was used in training the net and the remaining half was used in testing the predictive capacity (the generalization) of the net. When training, the data was sampled randomly. We used a step size $\eta = 0.01$ and a momentum parameter $\alpha = 0.9$ (corresponding to the procedure used by Odewahn et al.) About $40 - 150$ thousand passes were made through the training set and after each 200 passes, the net was examined for accuracy of learning and predictive capacity.

For the small data set, we used a $3, 5, 1$ network. The notation $3, 5, 1$ stands for three input neurons (corresponding to the input values for RF, SA, and AE), one hidden layer with five neurons, and one neuron in the output layer (for the output value of SR or BT). The input values were normalized to lie in $[0.25, 0.75]$. We tested four cases which we will denote by MM, ML, BL, and BB. The notation MM denotes modular neurons in both the hidden and output layers, and ML denotes modular neurons in the hidden layer and a linear neuron in the output layer. Likewise, BB stands for basic or standard neurons in both hidden and output layers, and BL for basic neurons in the hidden layer and a linear neuron in the output layer.

Figures 3 to 7 show our results. In each case, we show only the best cases for the modular and basic neurons, i.e., one case each from MM, ML and from BB, BL. In the figures $\delta_{train}$ and $\delta_{test}$ are the absolute values of the fractional discrepancy between the actual and the target output values of the network, averaged over the training set and the test set respectively.

For the surface roughness (SR) data shown in Fig. 3, the best performance was from the case MM, i.e., from using modular neurons in both the hidden and output layers. This choice made the energy and $\delta_{train}$ decrease fastest. On the test data, as shown in Fig. 3(c), the choice MM resulted in the best generalization. The minimum value for $\delta_{test}$ was already achieved after about 10,000 passes for MM, while for BL it was not yet achieved after 40,000 passes.

For the bore tolerance (BT) data shown in Fig. 4, the energy and training were best with ML. However, whereas ML achieved its best value for $\delta_{test}$ earlier than BL, the BL value was a little lower.

Figure 5 shows a 3,5,2 net used to compute both the SR and BT data simultaneously. Once again, the best case performances (MM and BL) are quite comparable.

We now turn to the larger data set from the star/galaxy discrimination study. The net we used here was 14,7,6,1. Figures 6 and 7 show the results of the study. The energy minimum (Fig. 7) is reached faster with the modular neuron and the learning rate is also marginally better [Figs. 6(a),(c)]. In the case of generalization, the modular neuron identified stars slightly better [Fig. 6(b)] while the basic neuron was slightly better at picking out galaxies [Fig. 6(d)].

In summary, we have made a detailed analysis of a new modular type of neuron which we believe is a viable alternative to the standard McCulloch-Pitts neuron for certain analog implementations of neural networks and for software simulations as well. We showed that for two different test data sets, the new neuron performed comparably to or (in the case of the SR data) better than the standard neuron, in agreement with theoretical expectations. Our study leaves open the question of whether there is any *a priori* characterization of data sets for which the modular neuron gives improved performance.

## Acknowledgement

S.L.A. wishes to thank J. Atick for a number of helpful conversations, and the authors thank S. Odewahn for supplying the data set used in the star/galaxy discrimination study. This work was supported in part by the Department of Energy under Grant #DE–FG02–90ER40542.

**Figure Captions**

Fig. 1 (a) Electronic implementation of the modular neuron of Eqs. (2a,b), with only one input shown explicitly. (The offset signal $u_j$ is equivalent to an input $v_0^{in} = u_j$ coupling with synapse strength $w_{j0} = -1$.) Use of single phase inputs corresponds to excitatory synapses with $w_{jk}$ real and positive.
(b) Two phase inputs which generate real synapses $w_{jk}$ which can have either positive or negative sign. (c) Three phase inputs which generate complex number synapses $w_{jk}$, corresponding to Eqs. (2a,c).

Fig. 2 (a) $v_1^{in}, v_2^{in}$, and $v^{out}$ values for XOR of Eq. (7a), together with the modular neuron realization (shaded band) by Eq. (7b).
(b) $v_1^{in}$, $v_2^{in}$, and $v^{out}$ values for XNOR of Eq. (7a), together with the modular neuron realization (shaded band) by Eq. (7c).
(c) $v_1^{in}$, $v_2^{in}$, and $v^{out}$ values for a four-point dichotomy in 2 dimensions which cannot be generated by a modular neuron. Since all three -point dichotomies in 2 dimensions can be generated, this implies $d_{VC} = 3$ for N=2, in agreement with Eq. (10c).

Fig. 3 Results for the surface roughness (SR) fit to the small data set, giving
(a) cost function or "energy",
(b) fractional discrepancy on training set, and
(c) fractional discrepancy on test set. The curves labeled MM are for modular neurons in the hidden and output layers; the curves labeled BL are for basic (or standard) neurons in the hidden layer and a linear neuron in the output layer.

Fig. 4 Results for the bore tolerance (BT) fit to the small data set, giving
(a) cost function,
(b) fractional discrepancy on training set, and
(c) fractional discrepancy on test set. The curves labeled ML are for modular neurons in the hidden layer and a linear neuron in the output layer; the curves labeled BL are for basic neurons in the hidden layer and a linear neuron in the output layer.

Fig. 5 Results for a combined fit to surface roughness (SR) and bore tolerance (BT) for the small data set, giving
(a) cost function,
(b) fractional discrepancy on training set, and
(c) fractional discrepancy on test set. The curves labeled MM are for modular neurons in the hidden and output layers; the curves labeled BL are for basic neurons in the hidden layer and a linear neuron in the output layer.

Fig. 6 Results for the star/galaxy discrimination study using the larger data set, showing
(a) fraction of stars correctly classified on the training set,
(b) fraction of stars correctly classified on the test set,
(c) fraction of galaxies correctly classified on the training set, (d) fraction of galaxies correctly classified on the test set. The curve BBB is for basic neurons in the hidden and output layers; the curve MML is for modular neurons in the hidden layers and a linear neuron in the output layer.

Fig. 7 The cost function for the star/galaxy discrimination study using the larger data set. The curve BBB is for basic neurons in the hidden and output layers; the curve MML is for modular neurons in the hidden layers and a linear neuron in the output layer.